\documentstyle[preprint,aps]{revtex}
 \begin{document}
\preprint{USACH/99/03 ,  DFTUZ/99-06}
\title{INSTANTON-LIKE EXCITATIONS IN 2D FERMIONIC FIELD THEORY}
\author{ J.L. Cort\'es $^1$\thanks{E-mail: cortes@leo.unizar.es}
J. Gamboa$^2$\thanks{E-mail: jgamboa@lauca.usach.cl}, 
I. Schmidt$^3$\thanks{E-mail: ischmidt@fis.utfsm.cl} and J. Zanelli$^{2,4}$
\thanks{E-mail: jz@cecs.cl}} 
\address{$^1$Departamento de F\'{\i}sica Te\'orica, Universidad de
Zaragoza, 
Zaragoza 50009, Spain\\ 
$^2$Departamento de F\'{\i}sica, Universidad de Santiago de Chile, 
Casilla 307, Santiago 2, Chile \\ 
$^3$ Departamento de F\'{\i}sica, Universidad T\'ecnica F. Santa Maria,
Valpara\'{\i}so, Chile \\ 
$^4$ Centro de Estudios Cient\'{\i}ficos de Santiago, Casilla 16443, 
Santiago, Chile}

\maketitle
\vskip 2.0cm

In this article we review some recent results previously reported in 
\cite{1,2}, where a new type of nonperturbative excitations was identified 
in the Thirring and Schwinger models. This effect is explicitly observed 
in the light-cone quantization using functional methods. The effect arises 
from the need to compactify the $x^-$ coordinate in the light-cone, which 
in effect changes the spacetime topology. 

Integrating out the right-handed fermions ($\psi_R$) yields an effective 
action for left-handed fermions ($\psi_L$) which contains excitations similar 
to abelian instantons produced by a composite of $\psi_L$. Right-handed 
fermions don't have a similar effective action and therefore, quantum 
mechanically, the symmetry $\psi_L \leftrightarrow \psi_R$ 
must be broken as a result of the topology. 
The conserved charge associated to the topological states is quantized. 
Different cases with only fermionic excitations or bosonic excitations or 
both can occur depending on the boundary conditions and the value of the 
coupling.

Alternatively, one can start from an action including an auxiliary vector 
field which, upon integration of $\psi_R$, carries instanton excitations 
\cite{2}. In what follows, we discuss the first approach and mention in 
the conclusions the differences with the second alternative.

In recent years quantization in the light cone frame has been extensively 
studied in connection with the discovery of new non-perturbative effects 
that would be unobservable in the standard spacetime quantization 
\cite{brodsky1,wilson,bigatti}.

There are several reasons that make the light cone quantization a method 
radically different compared to the standard quantization. The dispersion 
relation $k_\mu k^\mu = m^2$becomes $k^+k^- =m^2$, in the light cone. 
This implies that the particles and antiparticles occupy disconnected 
sectors of momentum space. Furthermore, if the momentum $k^+$ is 
discretized by compactifying $x^-$, the energy $k^-$ is also quantized 
and nonzero. Thus, the light cone momentum is never singular if the 
spacetime topology is $S^1\times \Re$ \cite{brodsky2}.

The above comments show that the quantization in the light cone is 
naturally defined over a manifold with non-trivial topology. As a 
consequence, one could expect new physical effects originated in the 
implicit difference in topology  from standard spacetime quantization.

 From this point of view, the massless Thirring model is an example where
one could investigate the new effects that emerge in the light-cone 
frame. We will show below that, contrary to the standard quantization,
the non-trivial topology of the light-cone spacetime induces abelian
instanton-like excitations, {\it i.e.} a purely quantum mechanical effect 
that appears in the calculation of the fermionic determinant.

Let us consider the massless Thirring model in the light cone frame

\begin{equation}
{\cal L} = i \psi^{\dagger}_L \partial_+ \psi_L + i \psi^{\dagger}_R
\partial_- \psi_R - 2g^2 (\psi^{\dagger}_L \psi_L)( \psi^{\dagger}_R
\psi_R),
\label{thirring1}
\end{equation}
where $\partial_\pm = \frac{\partial}{\partial x^\pm}$ and $x^+$ and $x^-$
play the role of time and space respectively. 

One should note that (\ref{thirring1}) is invariant under the charge  
conjugation symmetry $\psi_{L,R} \leftrightarrow \psi^{\dagger}_{L,R}$, 
that is expected to be conserved at the quantum level. However in order 
to quantize the system following the path integral methods it is more 
convenient to write (\ref{thirring1}) as follows
\begin{eqnarray}
{\cal L} &=& \psi^{\dagger}_R {(i\partial_- - 2g^2 \psi^{\dagger}_L
\psi_L)}\psi_R + \psi^{\dagger}_L {(i\partial_+) }\psi_L, \nonumber
\\
&=& \psi^{\dagger}_R {(i\partial_- + 2A_-)}\psi_R + \psi^{\dagger}_L
{(i\partial_+) }\psi_L. \label{3}
\end{eqnarray}
where $A_- = -g^2 \psi^{\dagger}_L \psi_L$.

The symmetry under charge conjugation including the auxiliary field 
$A_-$ is now 
\begin{equation}
\psi^{\dagger}_L \leftrightarrow \psi_L, \,\,\,\,\,\,\,\, \psi^{\dagger}_R
\leftrightarrow \psi_R, \,\,\,\,\,\,\, A_- \rightarrow -A_-. \label{4}
\end{equation}

Thus, the generating functional,
\begin{equation}
Z = \int {\cal D} \psi^{\dagger}_R {\cal D}\psi_R {\cal D}
\psi^{\dagger}_L {\cal D}\psi_L \,\,\, e^{i S}, \label{5}
\end{equation}
after integrating over the right handed fields is
\begin{equation}
Z = \int {\cal D} \psi^{\dagger}_L {\cal D}\psi_L  \det ( i \partial_- +
2A_- ) \,\,\,e^{i\int dx^+ dx^- \psi^{\dagger}_L i\partial_+ \psi_L},
\label{6}
\end{equation}
where $i \partial_- + 2A_- $ is a one-dimensional Dirac operator. It should 
be noted that  $A_- = A_-(x^-, x^+)$ is a function of two variables. The 
eigenvalues $\lambda_n$ of the equation
\begin{equation}
[i \partial_- + 2A_-] \varphi_n = \lambda_n \varphi_n, \label{7}
\end{equation}
are parametric functions of $x^+$, which can be 
determined integrating (\ref{7}) with $x^+$ fixed.

On the other hand, as $i \partial_- + 2A_- $ describes a fermionic system, 
usual practice would be to solve (\ref{6}) using antiperiodic boundary 
conditions. However, the periodic topology of $x^-$ also allows for twisted 
boundary conditions 
\begin{equation}
\psi (x^+, \sigma) = e^{2\pi i\gamma(x^+)} \psi (x^+, 0), \label{8}
\end{equation}    
where the real parameter $\gamma$ should be fixed by quantum consistency.

The solution of (\ref{7}) is 
\begin{equation}
\varphi_n (x^+, x^-) = e^{i \int_0^{x^-}
dy{[2A_-(x^+, y) - \lambda_n]}}\chi_0,\nonumber
\end{equation}
where $\chi_0$ is a Grassmann spinor independent of $x^-$ . Using (\ref{8}) 
the eigenvalues are found to be   
\begin{equation}
\lambda_n = - \frac{2\pi (\gamma + n)}{\sigma} + \frac{2}{\sigma}
\int_0^{\sigma} dx^- A_-. \label{9}
\end{equation}

Using $\zeta$-function regularization, the determinant is found to be 

\begin{equation}
\Gamma_\gamma (A) = \det ( i\partial_- + 2 A_- ) = {\cal N} 
\sin ( \Phi - \pi \gamma ),
\label{10}
\end{equation}
where $\Phi = \int_0^{\sigma} dx^- A_-$ is the \lq \lq magnetic flux"
produced along the surface
$x^+ = {\mbox const.}$ and ${\cal N}$ is a normalization constant.

Thus, the effective quantum theory for the left handed fields reads 
\begin{equation}
Z = \int \prod_{x^+}{\cal D}\psi^{\dagger}_L {\cal D}\psi_L\,\,\, 
\Gamma_\gamma (A) \,\,\,  e^{i\int dx^+ dx^- \psi^{\dagger}_L 
(i\partial_+)\psi_L }, \label{11}
\end{equation}
with $\Gamma_\gamma (A)$ given by (\ref{10}).

In order to preserve quantum mechanically the charge conjugation symmetry,
the boundary conditions have to be consistent with the transformation
$\psi^{\dagger} \leftrightarrow \psi$ . This requires that 
\begin{equation}
e^{2\pi i\gamma} = e^{- 2\pi i\gamma}
\label{cbc}
\end{equation}
and then $\gamma$ must be an integer (periodic boundary conditions)
or a half-integer (antiperiodic boundary conditions).
The next step is to observe that the boundary condition (\ref{8}) is 
invariant under the shift $\gamma \rightarrow \gamma + 1$. This implies 
\begin{equation} 
\Gamma_\gamma(A) =\Gamma_{\gamma+1}(A),
\label{112}
\end{equation}
which requires $\sin(\Phi -\pi\gamma) = \sin(\Phi -\pi(\gamma + 1))$, 
or equivalently 
\begin{equation}
\sin (\Phi - \pi \gamma ) = 0. \label{12}
\end{equation}
In view of the assertion (\ref{cbc}), the only consistent solutions 
of (\ref{12}) are:
\begin{equation}
\Phi = (m+ \frac{1}{2}) \pi, \,\,\,\,\, 
\gamma =m'+\frac{1}{2}, \,\,\,\,\mbox{with}\,\,\,\,\,\, 
m, m'\in\,\,\,\, \mbox{{\bf Z}},
\label{13}
\end{equation}
or,
\begin{equation}
\Phi = m \pi, \,\,\,\,\,\gamma =m', \,\,\,\,
\mbox{with}\,\,\,\,\,\,\, m, m'\in \mbox{{\bf Z}}.
\label{131}
\end{equation} 
In summary, the Thirring fields obey either antiperiodic boundary 
conditions with half-integer flux, or periodic boundary conditions 
with integer flux.

As a consequence of (\ref{12}) the fermionic determinant satisfies
the condition
\begin{equation}
\Gamma_\gamma (A) = \Gamma_{-\gamma} (- A).
\label{111}
\end{equation}
and the theory is invariant under charge conjugation.
In fact, condition (\ref{12}) means that the generating function for the 
effective theory vanishes. This should come as no surprise because, 
as noted by 't Hooft, the path integral for a massless fermion vanishes 
when the Fermi field couples to a gauge field with nontrivial 
topology \cite{thooft} (see also \cite{kiskis}). 

This result shows that there are two points of view to analyze this 
problem.  Before integrating out $\psi_R$ each Fermi field interacts 
with a background made out of fermions of the opposite chirality.  
After integrating out one of the two species, the remaining field 
self interacts with its own condensate $\psi^\dagger \psi$.  In our 
discussion $\psi^{\dagger}_L \psi_L$ plays the role of an external 
background gauge field $A_-$ with non-trivial topology interacting 
with the  $\psi^{\dagger}_L$.  As a consequence of the light-cone 
quantization, the left-right symmetry is broken because $x^-$ is 
compactified but not $x^+$. Thus, although the starting classical 
action (1) is symmetric under $\psi_L \leftrightarrow \psi_R$, quantum 
mechanically only left-handed charge is conserved.

In fact, the conserved Noether charge associated with a rigid phase 
rotation of the effective action (\ref{11}) is
\begin{equation}
Q_- =: \int_0^\sigma dx^- \psi^{\dagger}_L \psi_L ,
\label{Q}
\end{equation}
while a similar conserved charge for the right-handed fermions is not 
defined in the effective theory. Furthermore, this charge is quantized 
by virtue of (\ref{13}) or (\ref{131}).
In the previous form of the path integral (\ref{5}), the right-handed 
charge $Q_+=: \int_0^\sigma dx^- \psi^{\dagger}_R \psi_R $, is 
classically conserved but obeys an anomalous quantum algebra with $Q_-$.

At this point one could note that the expression for the determinant of 
the one-dimensional Dirac operator obtained by using $\zeta$-function 
is not the most general expression for the regularized determinant. The 
general form of the regularized one-dimensional determinant is 
\begin{equation}
\Gamma_\gamma (A) = e^{a + b\xi} \sin \xi, \label{regu}
\end{equation}
where $\xi = \Phi - \pi \gamma$ and $a$ and $b$ are arbitrary coefficients 
that depend on the regularization procedure. This general form 
is obtained starting from the formal expression of $\Gamma_\gamma (A)$
as an infinite product of eigenvalues depending on $\xi$. In order
to give meaning to this product one considers the logarithm as a 
(divergent) sum and takes enough derivatives with respect to $\xi$
(two) until a convergent sum, which is the series representation of
$\sin^{-2}\xi$, is obtained. The linear expression $a + b \xi$ gives 
the most general $\xi$-dependence due to the divergence of 
the one-dimensional determinant. 
The result of $\zeta$-function regularization corresponds to the particular 
choice of the regularization parameter $b = 0$ and $\,e^{\,a\,}\,$ is the 
normalization constant ${\cal N}$ in (\ref{10}).

If one takes the general form (\ref{regu}) for the regularized 
one-dimensional determinant, the consistency condition (\ref{112})
leads now to
\begin{equation}
\left[ 1 + e^{-b\pi} \right] \sin (\Phi - \pi \gamma ) = 0 
\label{12reg}
\end{equation}
instead of (\ref{12}), and this implies once more a quantization
of the flux, (\ref{13}) or (\ref{131}) , for $b\neq i(2n+1)$ .
The possibility to have a charge conjugation invariant theory
without a quantized flux by choosing appropiately the 
regularization, $b = i(2n+1)$ , deserves further investigation.
 
\noindent We end up this discussion by pointing out that an important 
information on the non-perturbative spectrum of the model can be read 
from the relation, $\Phi=-g^2Q_-$, between the quantized flux $\Phi$ 
and the conserved charge $Q_-$ which counts the number $n_L$ of 
left-handed fermions in a fixed-$x^+$ surface.

\noindent In the case of periodic boundary conditions $m\pi=-g^2n_L$;
then one has $m=n_L=0$, i.e. only neutral bosonic excitations, unless 
$g^2/\pi$ is a rational number. 
If $g^2/\pi=p/q$ then the quiral charge has to be a multiple of $q$ 
and the magnetic flux will be $\Phi=-\pi n_Lp/q$. For $q$ even one has 
bosonic excitations while for $q$ odd the spectrum contains both bosonic 
and fermionic excitations.

\noindent In the case of antiperiodic boundary conditions one has
$(m+1/2)\pi=-g^2n_L$ and $g^2/\pi$ has to be a rational number with 
even denominator. Only states, characterized by an integer $n$, with
a chiral charge $n_L=(2n+1)q/2$ and a flux $\Phi=-(n+1/2)p\pi$ appear
in the spectrum; for $q/2$ even one has a purely bosonic spectrum
but in the case $q/2$ odd only fermionic excitations are present.
  
\noindent The results presented here can be summarized as follows: (i) A 
fermionic topological  instanton-like excitation arising from the 
compactification of the $x^-$; (ii) Quantum mechanically, the massless 
Thirring model is analogous to the $\lambda \phi^4$ theory in 1+1 
dimensions, where an abelian instanton is also possible (see {\it e.g.} 
\cite{coleman}; (iii) different situations with only fermionic excitations, 
bosonic excitations or both can be identified ; (iv) the quantum 
system breaks the classical left-right symmetry.  

\noindent The lesson we've learned is that a simple self-interacting 
fermionic system can have many more excitations than those expected 
perturbatively.  This last statement is particularly interesting for 
$\mbox{QCD}_4$ at low energies where this kind of theory - {v.i.z.} the 
Nambu-Jona-Lasinio model- is expected. It remains to be proven whether 
a similar construction can be carried out in higher dimensions.

\noindent To end up let us point out that a quantization of the Thirring 
model in the light-cone compatible with the conservation of the vector
current is possible if an auxiliary vector field is introduced before 
integrating out the fermion fields. In this formulation the non-trivial 
topology of space-time, due to the compatification of the $x^-$ coordinate 
in the light-cone, leads to a quantization condition due to the boundary 
conditions. New excitations are also identified in this case as a consequence 
of the contribution of the zero modes. This formulation can be applied 
beyond the Thirring model; for an analysis of the excitations and vacuum 
structure of the Schwinger model and chiral Schwinger model see \cite{2}.

{\bf Acknowledgements}
We thank the organizers of the meeting for their kind hospitality in Buenos 
Aires. This work was partially supported by CICYT (Spain) project AEN-97-1680 
and grants 1980788, 1960229, 1960536, 7960001, and 7980045 from FONDECYT-Chile,
and DICYT-USACH. I.S. is a recipient of a C\'atedra Presidencial en 
Ciencias-Chile. Institutional support to CECS from Fuerza Aerea de Chile and 
a group of private companies (Business Design Associates, CGE, Codelco, Copec,
Empresas CMPC, Minera Colahuasi, Minera Escondida, Novagas, and Xerox-Chile), 
is also acknowledged.


\begin{references} 
\bibitem{1} J. L. Cort\'es, J. Gamboa, I. Schmidt and J. Zanelli, 
{\it Phys. Lett.} {\bf B 444} (1998) 451.
\bibitem{2}  J. L. Cort\'es and J. Gamboa, hep-th/990141,
{\it Phys. Rev.} {\bf D} (to be published).
\bibitem{brodsky1} S. Brodsky, H.-C. Pauli and S. Pinsky, hep-th/9705477, 
{\it Phys. Rep.} (in press).
\bibitem{wilson} M. M. Brisudova, R. Perry and K.G. Wilson, 
{\it Phys. Rev. Lett.} {\bf 78}, 1227 (1997); S. D. Glazek and K. G. Wilson, 
{\it Phys. Rev.} {\bf D57}, 3558 (1998).
\bibitem{bigatti} D. Bigatti and L. Susskind, hep-th/9711063.
\bibitem{brodsky2} T. Maskawa  and  K. Yamawaki, {\it Prog. Theor. Phys.}
{\bf 56} 270 (1976);
S. Brodsky and H. -C. Pauli, {\it Phys. Rev.}{\bf 32}, 
1993 (1985).
\bibitem{thooft} G. \rq t Hooft, {\it Phys. Rev. Lett.}{\bf 37}, 8 (1976);
{\it i.b.i.d.} {\it Phys. Rev.}{\bf D14}, 3432 (1976).
\bibitem{kiskis} J. Kiskis, {\it Phys. Rev.}{\bf 15D}, 2329 (1977).
\bibitem{coleman} S. Coleman, {\it Aspects of Symmetry}, Cambridge 
University Press; 
A. M. Polyakov, {\it Gauge Fields and Strings}, Harwood (1988).
\end{references}
\end{document}